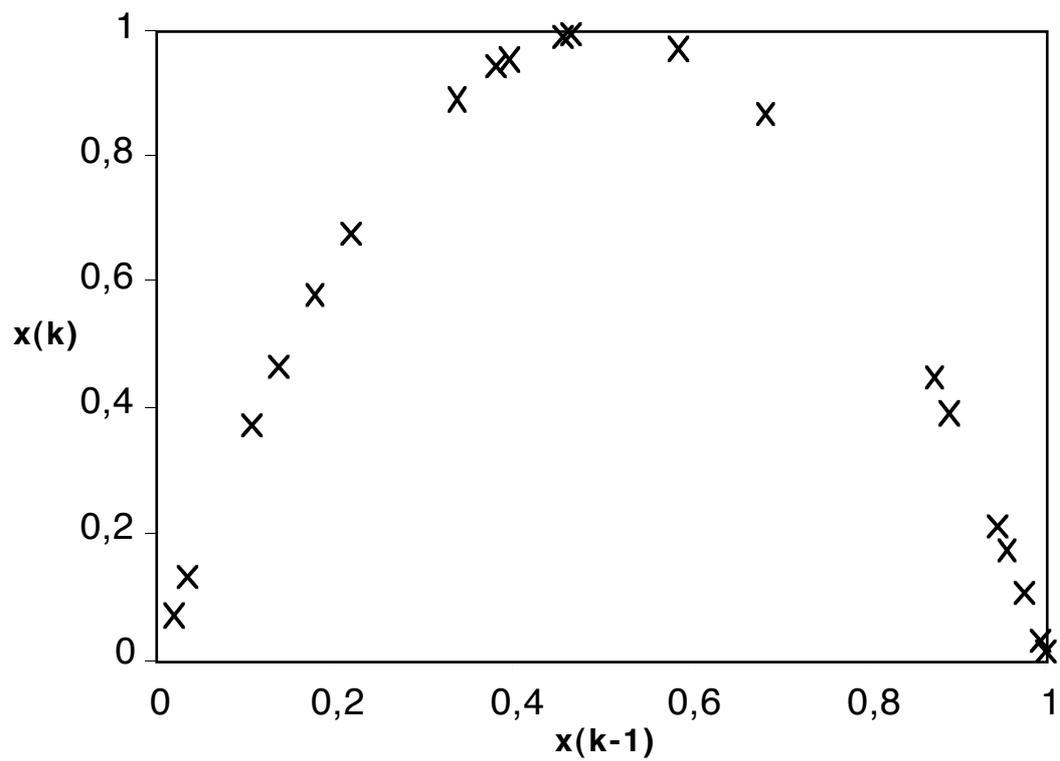

Figure 2
B.Lillekjendlie et. al.
Chaotic time series, Part II, ...
MIC

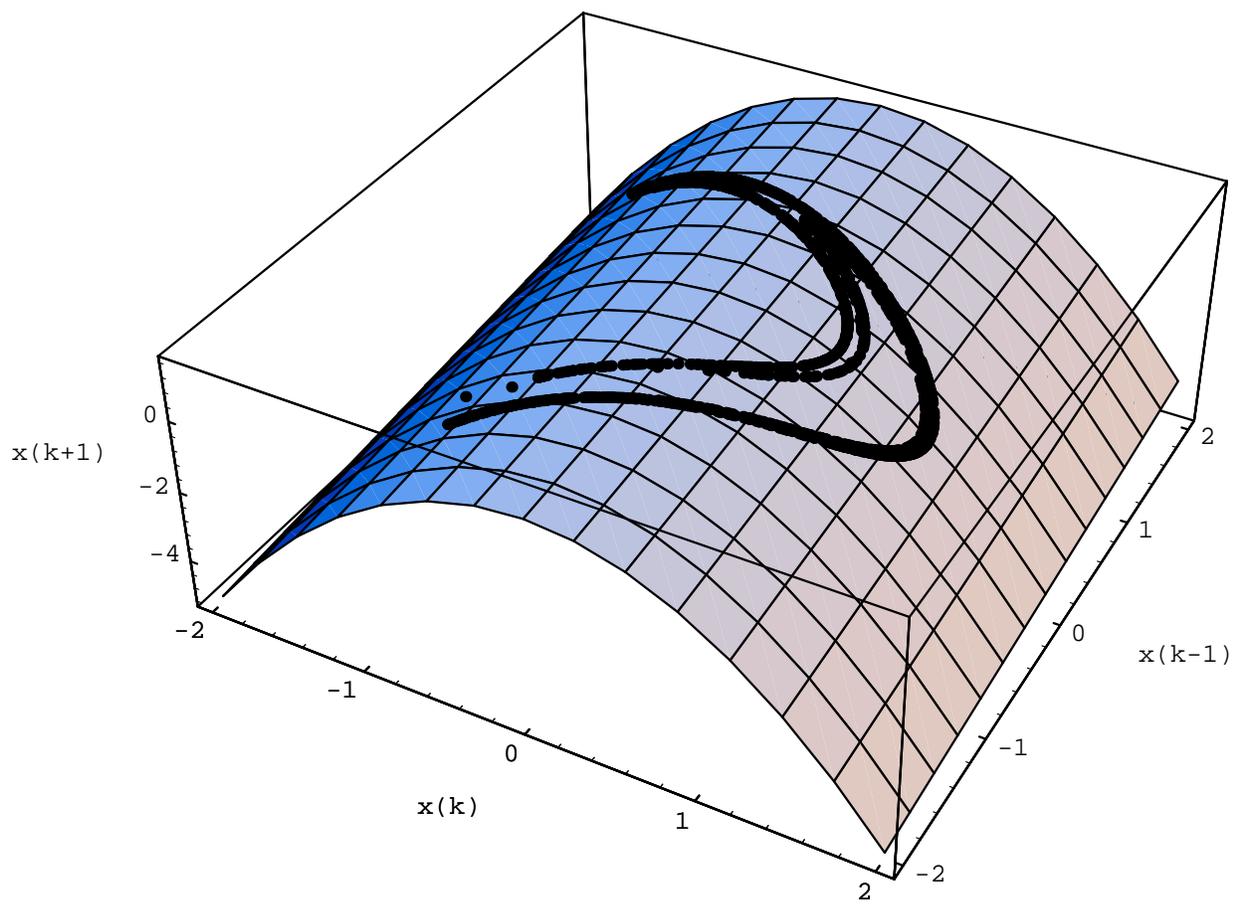

# Chaotic time series
# Part II: System identification and prediction

B. Lillekjendlie,* D. Kugiumtzis,† N. Christophersens †

January 14, 1994

**Abstract**

This paper is the second in a series of two, and describes the current state of the art in modelling and prediction of chaotic time series.

Sampled data from deterministic non-linear systems may look stochastic when analysed with linear methods. However, the deterministic structure may be uncovered and non-linear models constructed that allow improved prediction. We give the background for such methods from a geometrical point of view, and briefly describe the following types of methods: global polynomials, local polynomials, multi layer perceptrons and semi-local methods including radial basis functions. Some illustrative examples from known chaotic systems are presented, emphasising the increase in prediction error with time. We compare some of the algorithms with respect to prediction accuracy and storage requirements, and list applications of these methods to real data from widely different areas.

## 1 Introduction

The first paper in this series [30] discussed the phenomenon of chaotic behaviour, i.e. the fact that seemingly stochastic time series can be generated from low dimensional deterministic systems. Chaotic systems are characterised by features such as strange attractors and positive Lyapunov exponents, which, when estimated from real data, are used to identify chaos. From this starting point, the focus of the present paper is system identification and prediction; identification is also called the "inverse problem" in dynamical systems theory. Under ideal conditions, a chaotic and therefore randomly behaving system, may be identified using techniques from non-linear *deterministic* system identification. In some sense, a seemingly stochastic problem has a deterministic solution. However, in practice this is only partially true since the chaotic signal will often be corrupted by noise. But even with perfect reconstruction of the dynamical equations in the noise free case, only short term predictions are possible due to the extreme sensitivity of chaotic systems to uncertainties in initial conditions. This is because a chaotic system, while globally constrained to a finite region of state space, is locally unstable everywhere.

Prediction of chaotic time series is still a relatively new research topic, dating back to 1987. So far, we have identified more than 50 published articles in the field, with a steady growth. There may well be work that we are not aware of, and we apologise for any such omissions. Another review article, shorter than this paper, is [75].

Non-linear system identification and prediction at large is a diverse field with a plethora of potentially useful methods originating from different scientific disciplines. A broad set of these methods has been applied to chaotic systems and a survey in this area, which is the topic of the present paper, provides a useful comparison of different techniques. Hopefully, this will contribute both to spur the application of some of these methods to other areas of non-linear identification and prediction as well as providing useful feedback to the study of chaotic systems.

*Pb. 124, Blindern, N-0314 Oslo, Norway

†Department of Informatics,University of Oslo, Pb. 1080 Blindern, N-0316 Oslo, Norway

As noted in our first paper, one should also realise that even if chaos *per se* is not of interest, some knowledge of this phenomenon is highly useful in many non-linear studies. This is because chaotic behaviour is a pervasive non-linear phenomenon, and many systems may therefore turn chaotic in parts of the parameter space.

To limit the length of this article, we have chosen to omit system identification in the fields of fractals. Work like the Collage theorem by Barnsley [5] can be viewed as identification of system dynamics in the complex plane with applications mainly to data compression, whereas we focus on more conventional time series prediction.

In addition, noise in chaotic systems is not covered in this article, mainly because little theory has yet evolved in this area. For some initial work, see [24] and [34].

In the background section below, the notation as well as some basic mathematical concepts are given together with simple examples. The fundamental ideas behind each major approximation method class are then treated in sections covering global polynomials, multi layer neural network perceptrons, local polynomials, and semi-local methods, including radial basis functions. Here we give brief references to some applications. In the section 6 we discuss and compare the advantages of each method through tables and figures compiled from various sources.

## 2  Notation and mathematical background

For chaotic systems, delay coordinates are commonly used to generate state variables. We consider coordinates derived from scalar time series, but there are no principal difficulties in applying the same theory to multivariate observations. A general treatment of the so called embedding technique is given in [30]. For simplicity, in this article we let the delay time $\tau$ denote the fixed interval between observations (the sampling interval in the case of a continuous system), and consider the discrete scalar time series: $x_k = x(k\tau)$, where $k$ is an integer. The delay state vector at time $t = k\tau$ is defined as

$$\underline{x}_k = [x_k, x_{k-1}, \ldots, x_{k-(m-1)}]^T, \tag{1}$$

where $m$ is the embedding dimension and $T$ denotes the transpose. Note that the first element of the vector $\underline{x}_k$ is the sample value at time $k\tau$. The first complete state vector can be constructed at time $(m-1)\tau$ after the first sample arrives.

There are two equivalent ways of expressing the map from time $k\tau$ to $(k+1)\tau$,

$$\underline{x}_{k+1} = \underline{f}(\underline{x}_k) \qquad \underline{f} : \Re^m \mapsto \Re^m, \tag{2}$$
$$x_{k+1} = f(\underline{x}_k) \qquad f : \Re^m \mapsto \Re, \tag{3}$$

where the vector field $\underline{f}$ is related to scalar field $f$ as $\underline{f}(\underline{x}_k) = [f(\underline{x}_k), x_k, \ldots, x_{k-(m-2)}]^T$. For a chaotic system, one basically only knows that $f$ is non-linear. However, we will assume that $f$ is at least continuous and also continuously differentiable if needed. It is useful to note that geometrically equation (3) defines an $m$ dimensional surface (manifold) in $\Re^{m+1}$. By embedding space we mean $\Re^m$, and we denote the space $\Re^{m+1}$ where the surface exsits as the graph space.

If the time series stem from a chaotic system in its asymptotic state on a strange attractor of dimension $d$, Takensń theorem [68], and its extension [61], states that $m \geq 2d + 1$ components in the delay coordinate vector are sufficient to reconstruct the attractor for almost all dynamical systems. This implies that the vectors $\underline{x}_k$ all lie on the finite attractor in $\Re^m$ and that the observation pairs $(x_k, \underline{x}_{k-1})$, $(x_{k-1}, \underline{x}_{k-2})$, ..., $(x_2, \underline{x}_1)$ lie on the manifold generated by $f$ in $\Re^{m+1}$. The identification problem amounts to constructing an approximation $\hat{f}$ to $f$ given the observation pairs. This is a well known approach to system identification as outlined for instance in [36]. The problem of approximating a manifold in $\Re^{m+1}$ given points on, or near to its surface in the case of noise, is a central problem in numerical approximation theory as well as in statistical non-linear regression. Through Takensń theorem, identification of chaotic systems is put on a firm mathematical footing and shown to be a non-linear identification problem. In practice, the embedding dimension $m$ may be estimated by different methods and values less than $2d + 1$ may be feasible [30]. Choosing $m$ too large is not a problem in principle, but will certainly lead to a higher computational burden than necessary, and probably a less accurate predictor.

As a simple example, assume the well known logistic map $x_{k+1} = 4x_k(1 - x_k)$. A series of observations from this map looks like noise, and the autocorrelation of the data is as for white noise. Linear techniques will therefore be of no use in predicting such time series, but it is clear from the map itself that $m = 1$ will suffice to embed the attractor. In this case it is rather simple to build an approximation $\hat{f}$ to $f$ from the observation pairs $(x_k, x_{k-1})$, $(x_{k-1}, x_{k-2})$, ..., $(x_2, x_1)$ as illustrated in Fig. 1a). Here 20 pairs are plotted and the underlying shape of the one dimensional graph generated by $f$ in $\Re^2$ is clearly seen. If $m = 2$ was chosen, the result would be points on a one dimensional curve in $\Re^3$. As a second example, 200 observation pairs from the Henon map, which may expressed as $x_{k+1} = 1 - 1.4x_k^2 + 0.3x_{k-1}$ [26], are shown for $m = 2$ in Fig. 1b). The domain of variation in embedding space is the attractor which is recognised in the figure as lying on the surface in $\Re^3$ generated by $f(x, y) = 1 - 1.4x^2 + 0.3y$.

To assess $\hat{f}$, the normalised prediction error $e$ over a set of samples with $N$ elements is used:

$$e = \sigma_\delta / \sigma_x, \qquad (4)$$

where $\sigma_\delta$ is the root mean square prediction error given by $\sigma_\delta^2 = \frac{1}{N} \sum_i (x_i - \hat{f}(\underline{x}_{i-1}))^2$, and $\sigma_x$ is the sample standard deviation, $\sigma_x^2 = \frac{1}{N} \sum_i (x_i - \bar{x})^2$, where $\bar{x}$ denotes the average of the $x$ values. If $e \approx 0$, the prediction is almost perfect, whereas an $e$ value equal to 1 is equivalent to using the average as the predictor.

We will mainly think of identification in batch mode, where the model is built to minimise the sum of the root mean square error over all samples in a training set. Good statistical practice dictates that the prediction error $e$ of the estimated model should not be computed from samples used to construct $\hat{f}$, but over a separate test set. When few observation pairs are available, the standard technique is cross validation [66].

On-line applications are feasible when the methods described here are used to continuously update $\hat{f}$; think of this as an analog to a Kalman filter doing on-line model parameter estimation. A non-linear model estimated off-line can also be used in the predicting step of an on-line, extended Kalman filter.

Maps $\hat{f}$ approximating $f$ in (3) are one-step predictors. If it is desirable to predict more steps ahead in spite of the escalating uncertainty, say $r > 1$ steps, one can repeat the one-step prediction $\hat{f}$ $r$-times. Alternatively one may estimate the $r$-th iterate $x_{k+r} = f^{(r)}(\underline{x}_k) = f(\underline{f}(\cdots \underline{f}(\underline{x}_k)))$ directly. In [21] and [9] it is argued that iterated predictors outperform direct ones. Intuitively, when the prediction horizon $r$ increases, the function $f^{(r)}$ gets very complex and hard to approximate, which is illustrated in Fig. 2 showing the logistic map together with its 2nd, 4th, and 6th iterates. Direct approximation of $f$ over only a few time steps quickly becomes intractable because of the wild oscillations occurring.

In [1] and [2] it is shown that minimising the prediction error $\sigma_\delta$ often leads to a $\hat{f}$ which does not reproduce dynamic invariants in the original data like Lyapunov exponents and the density of points on the attractor. They suggest that the performance criterion should also include a fit to these invariants, and achieve this by not only predicting $x_{k+1}$ as $\hat{f}(\underline{x}_k)$, but as a linear combination of the $L$ first iterates $\hat{f}(\underline{x}_k)$, $\hat{f}(\hat{f}(\underline{x}_{k-1}))$, ..., $\hat{f}(\hat{f}(\cdots(\hat{f}(\underline{x}_{k-L}))$. Since high iterates defy approximation, $L$ should be low. A system identified this way shows slightly larger prediction errors, but will reproduce the general behaviour of the underlying system better.

An important property of chaotic systems facilitates the approximation task. A sufficiently long time series produces a sequence of vectors $\underline{x}_k$ that is approximately dense on the attractor, meaning that no new $\underline{x}_k$ will be far from those already observed. Thus, borrowing a term from the identification field, a chaotic system is in some sense "persistently excited". Further, a low dimensional attractor occupies only a fraction of the higher dimensional embedding space, which can reduce storage requirements in approximations significantly.

# 3 Global approximation methods

## 3.1 Global polynomials

An obvious approximation $\hat{f}$ to $f$ (or $\hat{\underline{f}}$), is a polynomial in the $m$ delay coordinate variables of degree $p$, set by the user. Polynomials can be written

$$\hat{f}(\underline{x}_k) = \sum_i w_i \phi_i(\underline{x}_k), \qquad (5)$$

where $w_i$ are parameters and the basis functions $\phi_i$ are powers and cross products of the components in $\underline{x}_k$. Small adjustments to the weights $w_i$ will cause the map to change almost everywhere, and therefore this method is classified as global.

Since the parameters enter linearly, they can be fitted to the data using standard least squares techniques involving the normal equations—in practice often done by singular value decomposition [54]. A potential advantage is that the parameters may be estimated recursively (and therefore in real time as measurements accumulate), using a Kalman filter like algorithm [63]. [23] showed an alternative, efficient way of estimating the parameters using orthonormal polynomials and assuming ergodicity. In that case the multivariable parameter estimation problem can be reduced to simple computations involving sums of powers of the variables in the delay coordinate vector.

The simplest first order polynomial approximator is the well known Auto Regressive (AR) model [55]:

$$\hat{f}(\underline{x}_k) = \sum_{i=0}^{m-1} w_i x_{k-i} + w_m, \qquad (6)$$

which geometrically amounts to fitting an $m$ dimensional hyperplane to the data in $\Re^{m+1}$, and is thus a global linear model. This is an unsuitable model for predicting chaotic time series, but some authors have used AR models as benchmarks with which to compare non-linear techniques. Among the work trying global polynomials for predicting chaotic time series, we mention [31], [21], [9], [10], [50], and [23]. We refere to section 6 for a discussion of the quality of such global polynomial approximators.

A disadvantage using polynomials is that the number of independent parameters equals $\binom{m+p}{m}$, which gets intractably large as $p$ increases. Many independent parameters also increase the risks of overfitting noisy time series and higher order polynomials may show strong oscillations between samples. In addition, some scalar fields that are not well approximated by polynomials and in such cases rational functions may be used. The reason is basically that rational functions may have poles. If the underlying function has poles, even in the complex extension, these poles may ruin real valued approximations by plain polynomials. Rational functions were tried in [10].

## 3.2 Multi layer perceptron neural networks

Another class of global methods which have been applied to chaotic time series is multi layer perceptron (MLP) neural networks. These have an elaborate structure with sigmoid shaped basis functions like for example $\phi(x) = \tanh(x)$ or $\phi(x) = 1/(1 + e^{-x})$, and are probably the most commonly used neural networks. The building blocks in a neural network are the "nodes", which is just one basis function with some preprocessing of the input, typically an inner product. As a simple example, an MLP net with two input variables, three hidden nodes and one output node defines a function from $\Re^2$ to $\Re$, as illustrated in Fig. 4. Such a net can be written

$$\hat{f}(\underline{x}_k) = \phi(\sum_{i=0}^{2} w_i \phi(w_{0,i} x_k + w_{1,i} x_{k-1})). \qquad (7)$$

The $w_i$ and $w_{j,i}$ are real valued parameters, denoted weights in neural net terminology. A full net may have any number of layers and any number of nodes in each layer. In contrast to global polynomials, the weights in MLPs do not enter linearly, so iterative parameter estimation is required. Deriving the values of the weights in the net is in most cases done by back propagation, which is a steepest descent search [58]. As with polynomials, such MLP nets may approximate any smooth function $f : \Re^m \mapsto \Re$ to any degree of accuracy, given enough sigmoid functions with accompanying weights. A standard proof is found in [14].

In [31], MLP nets were applied for the first time to predict chaotic time series, namely data obtained from the Mackey-Glass equation [40]. Among other work, we mention [19] who applied a standard MLP to the Lorenz attractor [37], and extended this work in [20] to cover data from a controlled fluid dynamics experiment as well as estimates of sea surface temperatures. [7] predicted corporate bankruptcy, [39] did prediction on van der Pols oscillator and the thalamic neuron. We would also like to mention the work of [72], [73] and [74].

Typical disadvantages of standard MLP nets are the long parameter estimation time and potential local minima. To improve this, [76] tested a variation of the back propagation parameter estimation algorithm with a momentum term to speed up convergence, and simulated annealing [29] to avoid local minima. They analysed various data sets, including the logistic map and real data from two biological systems. A similar approach is found in [56]. Another successful approach to fast estimation algorithms, is [17], which devised a hierarchical way of structuring and estimating the weights in a sigmoid MLP. They tested the system behaviour on the logistic map and the Rössler system [57], and reported an improvement in parameter optimisation time of approximately three orders of magnitude.

## 4  Local methods

One of the main disadvantages of global methods is that a new sample pair $(x_k, \underline{x}_{k-1})$ may change $\hat{f}$ everywhere. Local interpolation overcomes this drawback by utilising only a limited number, say $s$, of neighbouring samples. There are two major classes of local methods, those applying neighbour samples directly in the prediction, and those fitting a function locally to the neighbours basing the prediction on the estimated function.

The simplest way to predict $x_{k+1}$ from neighbour samples, is to identify the nearest neighbour to $\underline{x}_k$ in the embedding space $\Re^m$. We denote the nearest neighbour to $\underline{x}_k$ by $\underline{x}_{k(1)}$, and the next sample $x_{k(1)+1} = f(\underline{x}_{k(1)})$ is then known from the time series, and can be used as the predictor. This was suggested by [38], and is equivalent to building a look-up table of previous state mappings. In terms of the original time series, one finds the segment of length $m$ that is "most similar" to $x_k, x_{k-1}, \ldots, x_{k-(m-1)}$ and then uses the sample following that segment to predict $x_{k+1}$, in other words $\hat{f}(\underline{x}_k) = x_{k(1)+1}$). This is also termed the "analog method". In [28], the method is used on a number of simulated data sets to distinguish chaos from coloured noise.

An improvement is to take the $s$ nearest neighbours and use the average of their state mappings as the predictor. Another variant was suggested in [67]. They selected $s = m + 1$ (not necessarily closest) neighbours to form the smallest $m$-dimensional simplex circumscribing $\underline{x}_k$ in $\Re^m$. $\hat{f}(\underline{x}_k)$ is then computed as a weighted sum of the mapped simplex corners. Besides synthetic data, they experimented with time series from measles, chicken pox and diatom populations (plankton). In [13] and [12], the method of Sugihara and May was used to predict vertical ground movements of an active caldera in Italy. In [35], an alternative formulation of the same algorithm was tested on driven semiconductors and the Lorenz attractor. Yet another variation is found in [44] which applies Voronoi tessellation methods from computational geometry [53] to build linear patches (tiles) in the $m + 1$ dimensional graph space.

A common mathematical formulation for all these methods is

$$\hat{f}(\underline{x}_k) = \sum_{i=1}^{s} f(\underline{x}_k(i)) \, \phi(\|\underline{x}_k - \underline{x}_{k(i)}\|), \qquad (8)$$

where $\underline{x}_{k(i)}$ denotes the $i$-th closest vector to $\underline{x}_k$ in $\Re^m$, $s$ is the number of neighbours, and $\|\cdot\|$ denotes a norm. Usually $\phi$ is a weight function increasing from zero to one when $\underline{x}_k$ approaches $\underline{x}_{k(i)}$. Note that there is no parameter estimation involved here, $\phi$ is a fixed function. Thus, this method is efficient in terms of computation time. However, the approximating maps are generally not continuously differentiable, and the search for neighbours become more time consuming as more vectors are stored.

The other class of methods fits a surface in graph space $\Re^{m+1}$, as described in section 2, to the measurement points $(x_{k(i)+1}, \underline{x}_{k(i)})$, $i = 1, 2 \ldots, s$. This may be a plane, but polynomials of higher degrees may also be used to interpolate between neighbours. Taking $s > m+1$ and fitting a plane, one obtains a local AR model, also called a local linear model. For chaotic time series, this was, as far as we know, first done in [21]. They experimented with such local AR models, as well as with higher order polynomials, but did not observe significant improvements moving to higher order. For comparison, they also applied a global AR model as a "standard forecasting technique". [10] continued to explore the relation between global and local AR models. Other work applying local methods is [9], [69], [50], [71], and [27]. Various versions of these techniques are well known in system identification, see for example [70].

## 5 Semi-local methods

Semi-local methods may combine the best of two worlds, the smoothness of global predictors and the localised dependence on new information of local predictors. Well known classes of semi-local approximators are splines and radial basis functions (RBF). For radial basis functions, three research communities exist. Approximation theorists are concerned with topics like convergence properties, see e.g. [52], people in the neural network community approach the problem from a more algorithmic point of view, see e.g. [33], and statisticians have their well developed field of kernel estimators, as described in e.g. [62]. To our knowledge, prediction of chaotic time series has only been considered from the standpoint of neural networks.

RBF-approximation can be thought of as a combination of the fitting and weighting approaches described in the previous section on local methods—weights are assigned according to the distance from the basis function centres, but these weights adjust parameters, not the next sample value as in (8). Applying $s$ basis functions, such approximators take the form

$$\hat{f}(\underline{x}_k) = \sum_{i=1}^{s} w_i \, \phi_i(\|\underline{x}_k - \xi^i\|), \tag{9}$$

where the function $\phi_i(r)$ is radially symmetric around a centre value $\xi^i$ in $\Re^m$, and $w_i$ are weights to be chosen. If $w_i$ are the only parameters to estimate, the normal equations can be utilised. If, however, there are parameters inside $\phi_i$ which enter the problem nonlinearly, time consuming iterative optimisation methods must be applied. As an intuitive visualisation of how radial basis functions work, consider Fig. 4 and imagine the smooth step functions replaced by, for example, Gaussian hats, (and a linear combination at the last level).

There are two main areas for modifying the basic scheme: experiments with various types of basis functions $\phi_i$ (section 5.1), and experiments with various algorithms for determining the parameters, especially the centres $\xi$ and the number $s$ of basis functions (section 5.2).

### 5.1 Choosing the basis function type

Various results prove that different types of RBF functions $\phi_i$ are universal approximators, that is, any smooth function can be approximated to any degree of accuracy given enough basis functions. This, in turns, requires an infinite amount of noise free measurement data. We refer to [49] for a neural network approach, [62] for a statistical approach and [45] for a result from approximation theory.

The most popular type is rotation symmetric Gaussian hats of the form $\phi_i(r) = \exp(-r^2/2\sigma_i^2)$, where the hat widths $\sigma_i$ are fixed constants. Even though Gaussian hats are global in theory, they decrease fast enough to have finite support for all practical purposes. The basis function can also be non-local like the multi quadric $\phi_i(r) = (r^2 + \alpha^2)^{-1/2}$, $\alpha$ and $r$ real. For a more extensive list of basis functions, see [8], and cf. [22] for a number of related methods viewed from numerical mathematics.

In higher dimensions, the locality advantage of Gaussian hats turns into a disadvantage. The basis functions are local, and since data is almost always scarce in higher dimension, most points in state space has no basis function cover. Another source of difficulty is that some input variables $x_{k-j}$ may be almost uncorrelated with the output variable $x_k$, especially if the embedding dimension $m$ is estimated too large. There are two typical ways to improve this situation, either by letting the hats smear out in some directions and become Gaussian "ridges", or by normalising the basis functions. The normalised Gaussian hats going into (9) are written

$$\phi_i(||\underline{x}_k - \xi^i||) = \frac{\exp(-||\underline{x}_k - \xi^i||^2/2\sigma_i^2)}{\sum_{j=1}^{s} \exp(-||\underline{x}_k - \xi^j||^2/2\sigma_j^2)}, \tag{10}$$

and they are named weight constant predictors (WCP) in [65] and [64]. They also suggested a method called weighted linear predictors (WLP) where the simple weights $w_i$ in (9) are replaced by the linear term $v_i + (\underline{x}_k - \xi^i) \cdot \underline{d}_i$ where $v_i$ is a scalar parameter and $\underline{d}_i$ is a parameter vector. In [42] and [43], the same method was tested on the Mackey-Glass equation, and in [1] and [2], a slightly modified weighting function was used. The WLP method was applied in [50] with $e^{-K_i}$ as the weight function, and neighbour samples taken from identified unstable periodic orbits (UPO) close to $\underline{x}_k$. Generic strange attractors can be approximated by unstable periodic orbits of a given length, and methods to identify such orbits are given in the same article. $K_i$ was the sum of the positive Lyapunov exponents of each neighbour UPO$_i$.

In [25], ridge functions were applied to the Mackey-Glass equation. An alternative view was proposed by [15], observing that a multi-dimensional Gaussian is equal to a product of scalar Gaussian with adjustable centres and widths. During learning, hat parameters are adjusted when a suitable Gaussian covers the input, otherwise a new hat is generated. In the product they use as few terms as possible, and new variables are introduced only when required, thus converting ridges into hats one variable at the time. In this way, the system identification method itself decides which previous variables that are sufficiently correlated to warrant inclusion into the model, estimation of the best embedding dimension $m$ becomes an integral part of the model building algorithm. This algorithm is named "CC-RAN". Another related tree-building algorithm was devised in [60] and [59].

## 5.2 The number of basis functions and their centres

In the most basic RBF method, there is one basis function at each sample so that $\xi^i = \underline{x}_i$. To reduce the computational burden, only the $s$ nearest samples are often taken into account. Now, $\xi^i = \underline{x}_k(i)$ and the number of terms in (9) is reduced. This approach was taken in [9], using non-local basis functions $r^3$ and the $s = 50$ nearest samples. The standard RBF method interpolates the data, and is thus sensitive to noise. To reduce this problem, Broomhead and Lowe [6] put a fixed number of basis functions on a regular grid. By letting the number of basis functions be less than the number of samples, the data was smoothed. They applied this method to predict the logistic map.

In higher dimensions, regular grids become infeasible because the number of basis functions grows exponentially with the dimension of the grid. In addition a chaotic attractor occupies only a small subset of the entire space, so most of the basis functions are superfluous. The solution is to represent only those parts of the embedding space $\Re^m$ where data exists, that is the manifold occupied by the attractor of the system. Thus, the memory requirements can be made proportional to the attractor's size, which also improves model estimation time and noise robustness.

One such algorithm, here denoted "hashing RBF", is described in [46]. Basis functions are maintained on a regular grid with spacing $\lambda$, but are only constructed at those grid points where data exists. A grid coordinate is related to the corresponding function parameters through a hash table, and once the neighbours are found, we are back to a summation model $\hat{f}_\lambda(\underline{x}_k)$ on the form (9). This hash table scheme was originally invented by Albus [4] and [3] for real time motion control in robotics. This is possible since a hash table makes the look-up extremely fast. Without detailed apriori information of the distribution of data in the input space, it is difficult to choose an optimal lattice spacing—a coarse grid will smooth data well, whereas a fine grid will capture details. A hierarchy of hash models spaced on grids with increasing resolution $\lambda$, written $\hat{f}(\underline{x}_k) = \sum_\lambda \hat{f}_\lambda(\underline{x}_k)$, will thus represent the major function structure on the coarsest scale grids, and add model details at finer grid resolutions.

Another idea for reducing the number of basis functions, is to cluster neighbouring sample vectors and represent them all with one basis function at the cluster centre $\xi^i$. When all training samples are collected, the clusters may be formed with for example the $k$-means clustering algorithm [41]. In [47] and [48], they applied a variation of this $k$-means clustering algorithm to centre the clusters, used the average distance the the neighbouring hats as hat width, and estimated the weights $w_i$.

The cluster centres and widths may also be built as an integral part of the parameter estimation, and one of the first descriptions of such algorithms was [32]. These algorithms add a new basis function only when no existing function covers the new sample, or if the existing functions cannot easily be changed to approximate the new sample. As more samples arrive, the hat widths are decreased gradually, leading to increased estimation accuracy. Usually such methods call for iterative estimation algorithms. In [51] a version of this algorithm was implemented, named "RAN" as an acronyme for resource allocating nets.

The automatic addition of basis functions during model identification represents one solution to the problem of so called model structure selection or model realisation. In statistics, there is an emerging theory of how to select the number of basis function, with cross validation and bootstrap methods as some of the themes [18]. Another challenge lies in selecting the form of the basis functions themselves, as well as identifying which variables that should go into the model. This is different from estimating the parameters in a fixed structure, which is usually what identification amounts to. Realisation theory as described in [11] is well developed for deterministic linear systems, but a fundamental theory lacks for non-linear systems (cf. [16]).

# 6 Discussions and comparisons

At this stage, a set of rules recommending certain methods for certain classes of problems, would be desirable. Unfortunately, such general advice is yet unavailable—there are too little experience gained from actual use. Instead, we have settled for the more modest goal of comparing experimental results given in the literature by compiling tables and figures. The figures and tables will hopefully give a feeling for the performance one can expect from the various methods.

The quality of the constructed approximators depends on many factors including: the underlying system, the number of samples available, the embedding dimension, the noise in process and measurements, the kind of approximator, the number of parameters in the approximator, the amount of computer and human resources invested in constructing the approximator, and the prediction interval. Based on results reported in literature, we have picked interesting experimental results where only one, or just a few, of these factors vary, the remaining factors remaining "relatively" fixed. The following tables and figures have been collated:

- Table (1) hints at how well different methods approximate the Mackey-Glass delay differential equation.

- Table (2) compares the approximation error for local AR models (local linear polynomials) with global AR models for different systems.

- Figure (3) shows how fast the prediction error increases with time for different methods and different test sets.

- Figure (5) shows how prediction accuracy is connected with the number of parameters and the number of samples for some methods applied to the Mackey-Glass equation.

Even though we have attempted to extract experimental results which are as comparable as possible, comparisons like these can never be completely fair, and only show the applicability of each method to the type of data chosen.

In all tables and figures, prediction error means normalised root mean square prediction errors as given in (4).

To give an impression of how well the various methods approximate a fixed system, we have in Table 1 collected prediction errors from experiments on the standard noise free Mackey-Glass equation with delay parameter $\Delta = 17$, with 500 training sets, embedding dimension $m = 4$, and sampling interval $\tau = 6$ time units. The difference Mackey-Glass eqution is written

$$x_{k+1} = \frac{0.2 x_{k-\Delta}}{1 + [x_{k-\Delta}]^{10}} - 0.1 x_k \,.$$

Most of the results are reported by the originators of the methods, presumably assuring maximum performance. Different number of samples were used in the test sets, typically either 500 or 1000. The number of model parameters differed between the tests. This is reasonable since each method should be allowed to apply an optimal number of parameters. Unfortunately, some reported prediction error 6 time units ahead, others 84, so the table reports results for both prediction intervals. In spite of this, the table should give an indication of how well the different methods approximate this Mackey-Glass system.

As can be seen from Table 1, no single method excels for one-step prediction 6 time units ahead, but global rationals and the weighted constant map (WCP) give the worst predictions. For 14 iterated one-step predictions (84 time units), multi layer perceptrons, local AR models, hashing RBF, weighted linear predictors (WLP), and CC-RAN all give prediction errors of approximately the same size. The only "bad" approximators here where the method of analogy and standard RBF.

Note that this table indicates that the prediction error increases with increasing prediction interval, which is quite natural. To illustrate this point further, we have in Fig. 3 collected prediction error as a function of prediction interval for three different systems and different system identification methods. The three systems are the Mackey-Glass equation, the Rayleigh-Bénard convection, and the Rössler equation. Again, the experimental conditions are similar enough to compare the results.

In Fig. 3, curves for the the Mackey-Glass equation with delay parameter $\Delta = 30$ are reproduced from [31]. The curves A), B), C), and D) are iterated and non-iterated global polynomials, and non-iterated and iterated MLP, respectively. These curves show how the prediction error increases with prediction time, and how non-linear methods may increase the prediction interval. Further, it can be seen that the iterated MLP predictor outperforms the direct long-time MLP predictor.

The Rayleigh-Bénard part of the same figure is based on [21], which used convection in an $^3$He $-$ $^4$He mixture with Rayleigh number $R/R_c = 12.24$ and fractal dimension $\approx 3.1$. Curve A) is global linear model with embedding dimension $m = 15$ (i.e. 15+1 parameters). Curves B), C), and D) are local linear methods with embedding dimension 4, 6 and 15 respectively; all predictors iterated. The improvement of local AR models over the global one should be obvious from this figure.

In the third part, we reproduce results for predicting the Rössler equation. The embedding dimension was 4, the sampling interval 0.87 time units, and 100 samples were used to train the hierarchical MLP net. The curve A) is from a global linear model, B) a local linear model, C) from global universal periodic orbits (UPO), D) from local UPO, and E) a global MLP model. Curves A) to D) are taken from [50], and E) from [17]. For this system, both global AR, but also local AR models give predictions which soon become totally unreliable, the universal periodic orbits give better predictions, but the MLP net is the most accurate approximator in this case.

Instead of varying the approximation method keeping most of the other factors constant, it is interesting to apply a small number of methods to a variety of processes, both simulated, laboratory experiments and real life data sets. Such a table can be compiled from [10]. In his article, local AR models are computed for all degrees of "locality" varying from the method of analogy to a global AR model. We have in Table 2 chosen the optimal local AR model and compared it with the global AR model estimated. The results can briefly be summarised as follows: With much noise and/or high dimensional systems, global AR models seem to outperform the local AR models; otherwise the local linear models are more accurate.

On noise free data, most of the approximation methods are, at least in principle, able to approximate any reasonably well behaved system to any degree of accuracy, given enough data. The approximation error will therefore depend on the number of samples used in the approximation process. For most methods, there is a strong relation between the number of samples used in training, and the number of parameters/weights in the model. We have therefore collated two graphs showing these relations in Fig. 5. All these curves are based on experiments with the standard Mackey-Glass equation as described above, with delay parameter $\Delta = 17$. The first graph shows prediction error as a function of the number of training samples, the second the prediction error as a function of the number of parameters. In both graphs, A) is the CC-RAN method [15], B) is adaptive clustering RBF [51], C) is hashing RBF [46], D) is K-means RBF [48] and E) is Standard RBF [47]. As can be seen, there are no major differences between most of the methods, except K-means RBF which requires far more training sets. Concerning the number of parameters against prediction error, it is clear that standard and K-means RBF requires lot of parameters compared to the other methods.

Since much of the discussions in this section has circled around deterministic, simulated systems, we conclude this section with Table 3 which summarises some of the real data sets analysed. We have not been able find a reasonable way of comparing these data sets, so the reader is referred to the original work.

# 7 Conclusion

Prediction of chaotic time series is a fairly new research topic, dating back to 1987. The underlying philosophy is geometrical, fitting non-linear functions to samples in embedding space $\Re^{m+1}$. The area of chaotic prediction is still relatively small within the larger domain of non-linear system identification and prediction, which offers a multitude of approaches still not tested on chaotic systems. Within non-linear studies at large, some knowledge of chaos is desirable since such behaviour is a pervasive non-linear phenomenon that may be manifest in various models in certain regions of the parameter space.

The fact that chaotic systems are "persistently excited" in the sense that accumulating points become dense on the attractor, is an advantage when modelling chaotic series. In particular, this could hold promise both for the local methods and the adaptive semi-local methods where the data determine the location and shape of the basis functions.

The paper has focused on methods that up to now have been applied to chaotic time series, including global polynomials, local polynomials, multi layer perceptrons and semi-local methods. Chaotic time series frequently resemble white noise having broadband Fourier spectra, and the non-linear predictors clearly outperform the standard linear methods like global AR models if the noise level is limited and the dimension of the attractor is low. The local and semi-local methods generally seem to be the best, but no non-linear method ranks top in all situations.

Most of the schemes can approximate any well behaved function to any desired accuracy level, provided enough samples and basis functions/parameters are available. The crucial question is therefore how many samples and parameters that are required to achieve a certain accuracy. From this point of view, the adaptive semi-local methods generally seem to be the best.

A crucial aspect is the robustness of the approximation schemes to noise. With noisy data we have, in the language of numerical mathematicians, a fitting problem and not an approximation problem. If the model is too "small", it will underfit the data and ignore important character-istics. If, on the other hand, the model is too "large", it will overfit the data, reproducing the noise as well as the underlying behaviour. It is impossible to use the number of parameters to measure the risk of overfitting, since in most methods the parameters are internally correlated. Currently, cross validation is the standard tool for selection of an appropriate model avoiding overfitting. However, the recent appearance of non-linear filtering schemes for time series allow for pretreatment of the data before constructing the model. This is certainly an interesting future line of development, since up to now, too little work has been done on noisy chaotic time series.

ACKNOWLEDGEMENTS

Support for this research was provided by the Norwegian Research Council (NFR) for D. Kugiumtzis and B. Lillekjendlie. B. Lillekjendlie was also supported by SINTEF-SI. The authors would also like to thank Tom Kavli and Erik Weyer, both at SINTEF-SI, for valuable discussions and careful reading of the manuscript.

# References

[1] Henry D.I. Abarbanel, Reggie Brown, and James B. Kadtke. Prediction and system identification in chaotic nonlinear systems: Time series with broadband spectra. *Physics Letters A*, 138(8):401–408, Jul 1989.

[2] Henry D.I. Abarbanel, Reggie Brown, and James B. Kadtke. Prediction in chaotic nonlinear systems: Methods for time series with broadband fourier spectra. *Physics Rewiev A*, 41(4):1782–1807, Feb 1990.

[3] James Albus. Data storage in the cerebellar model articulation controller (cmac). *Journal of Dynamic Systems, Measurement, and Control, Transactions of the ASME*, pages 228–233, Sep 1975.

[4] James Albus. A new approach to manipulator control: The cerebellar model articulation controller (CMAC). *Journal of Dynamic Systems, Measurement, and Control, Transactions of the ASME*, pages 220–227, Sep 1975.

[5] Michael Barnsley. *Fractals everywhere*. Academic Press, Inc., New York, 1988.

[6] D.S. Broomhead and D. Lowe. Multivariable functional interpolation and adaptive networks. *Complex Systems*, 2:321–355, 1988.

[7] David T. Cadden. Neural networks and the mathematics of chaos - an investigation of these methodologies as accurate predictors of corporate bankruptcy. In *Proc. of The First International Conference on Artificial Intelligence Applications on Wall Street*, New York, Oct 1991.

[8] Mats Carlin. Neural nets for empirical modelling. Master's thesis, Norwegian Institute of Technology (NTH), 1991.

[9] Martin Casdagli. Nonlinear prediction of chaotic time series. *Physica D*, 35:335–356, 1989.

[10] Martin Casdagli. Chaos and deterministic versus stochastic and non-linear modelling. *J. R. Statist. Soc. B*, 54(2):303–328, 1991.

[11] J.L. Casti. *Dynamical systems and their applications - Linear theory*. Academic Press, New York, 1977.

[12] M. Cortini and C.C. Barton. Nonlinear forecasting analyses of inflation-deflation patterns of an active caldera (campi flegrei, italy). *Geology*, 21:239–242, Mar 1993.

[13] M. Cortini, L. Cilento, and A. Rullo. Vertical ground movements in the campi flegrei caldera as a chaotic dynamic phenomenon. *Journal of Vulcanology and Geothermal Research*, 48:199–222, 1991.

[14] G. Cybenko. Approximation by superpositions of a sigmoidal function. *Math. Control, Signals Sys.*, 2:303–314, 1989.

[15] Gustavo Deco and Jürgen Ebmeyer. Coarse coding resource-allocating networks. *Neural Computation*, 5(1):105–114, Jan 1993.

[16] James R. Deller. Set membership identification in digital signal processing. *IEEE ASAP Magazine*, 4:4–20, 1989.

[17] J. Deppisch, H.U. Bauser, and T. Geisel. Hierarchical training of neural networks and prediction of chaotic time series. *Physics Letters*, A(158):57–63, 1991.

[18] Bradley Efron. *The Jackknife, the Bootstrap and Other Resampling Plans*. Society of Industrial and Applied Mathematics, 1982.

[19] J.B. Elsner. Predicting time series using a neural network as a method of distinguishing chaos from noise. *J. Phys. A: Math. Gen*, A(25):843–850, 1992.

[20] J.B. Elsner and A.A. Tsonis. Nonlinear prediciting, chaos and noise. *Bullerin American Meteorological Society*, 73:49–60, Jan 1992.

[21] J.D. Farmer and J.J. Sidorowitch. Predicting chaotic time series. *Physical Review Letters*, 59(8):845–848, Aug 1987.

[22] Richard Franke. Scattered data interpolation: Tests of some methods. *Mathematics of Computation*, 38(157):181–200, Jan 1982.

[23] M. Giona, F. Lentini, and V. Cimagalli. Functional reconstruction and local prediction of chaotic time series. *Pysical Rewiev A*, 44(6):3496–3502, Sep 1991.

[24] Peter Grassberer, Thomas Schreiber, and Carsten Schaffrath. Non-linear time series analysis. *Int. J. on Bifurcations and Chaos*, 1:521, 1991.

[25] Eric Hartman and James D. Keeler. Predicting the future: Advantages of semilocal units. *Neural Computation*, 3(3), 1991.

[26] M. Henon. A two dimensional mapping with a strange attractor. *Comm. Math. Phys.*, 50:69–77, 1976.

[27] Jr. Hunter, Norman F. Nonlinear prediction of speach signals. In Martin Casdagli and Stephen Eubanks, editors, *Nonlinear Modelling and Forecasting*, pages 467–492. Addison-Wesley, 1992.

[28] M.B. Kennel and S. Isabelle. Method to distinguish possible chaos from colored noise and to determine embedding parameters. *Physical Review A*, 46(6):3111–3118, Sep 1992.

[29] S. Kirkpatrick, C.D. Gelatt Jr., and M.P. Vecchi. Optimization by simulated annealing. *Science*, 220(4598):671–680, May 1983.

[30] D. Kugiumtzis, B. Lillekjendlie, and N. Christophersen. Chaotic time series, part i, estimating. *Modelling, Identification and Control*, 1993.

[31] Alan Lapedes and Robert Farber. How neural nets work. In D.Z. Anderson, editor, *Neural Information Processing Systems*, pages 442–456. American Institute of Physics, New York, 1987.

[32] Sukhan Lee and Rhee M. Kil. Multilayer feedforward potential function network. In *IEEE International Conference on Neural Networks*, pages I–161 – I–171, San Diego, 1988.

[33] Sukhan Lee and Rhee M. Kil. A gaussian potential function network with hierarchically self-organizing learning. *Neural Networks*, 4:207–224, 1991.

[34] A.J. Lichtenberg and M.A. Lieberman. *Regular and chaotic dynamics*. Springer-Verlag, 2nd edition, 1992.

[35] Paul S. Linsay. An efficient method of forecasting chaotic time series using linear interpolation. *Physics Letters A*, 153(6,7):353–356, Mar 1991.

[36] Lennart Ljung. Issues in system identification. *IEEE Control Systems*, pages 25–29, Jan 1991.

[37] E.N. Lorenz. Deterministic non-periodic flows. *Journ. of Atmospheric Science*, 20:130–141, 1963.

[38] E.N. Lorenz. Atmospheric predictability as revealed by naturally occuring analogies. *Journ. of Atmospheric Science*, 26:636, 1969.

[39] D. Lowe and A.R. Webb. Time series prediction by adaptive networks: a dynamical system perspective. *IEE Proceedings-F*, 138(1):17–24, Feb 1991.

[40] M. Mackey and L. Glass. Oscillation and chaos in physological control systems. *Science*, 197(287), 1977.

[41] J. MacQueen. Some methods for classification and analyses of multivariate observations. In L.M. LeCam and J. Neyman, editors, *Proc. of the 5th Berkeley Symp. on Mathematics, Statistics and Probability*, 1967.

[42] W.C. Mead, R.D. Jones, Y.C. Lee, C.W. Barnes, G.W. Flake, L.A. Lee, and M.K. O'Rourke. Using cnls-net to predict the mackey-glass chaotic time series. In *Proc. IEEE Int. Joint Conf. on Neural Networks (IJCNN)*, pages II485–II490, New York, 1991. IEEE.

[43] W.C. Mead, R.D. Jones, Y.C. Lee, C.W. Barnes, G.W. Flake, L.A. Lee, and M.K. O'Rourke. Prediction of chaotic time seires using cnls-net, example: The mackey-glass equation. In M. Casdagli and S. Eubank, editors, *Nonlinear Modelling and Forecasting*, pages 3–24. Addison-Wesley, 1992.

[44] A. I. Mees. Tesselation and dynamical systems. In M. Casdagli and S. Eubank, editors, *Nonlinear Modelling and Forecasting*, pages 3–24. Addison-Wesley, 1992.

[45] C. A. Michelli. Interpoaltion of scattered data: distance matrixes and conditionally positive definite functions. *Contructive Approximation*, 2:11, 1986.

[46] John Moody. Fast learning in multi-resolution hierarchies. In D.S. Touretzky, editor, *Advances in Neural Information Processing Systems 1*. Morgan Kauffman, 1989.

[47] John Moody and Cristian J. Darken. Learning with localized receptive fields. In Touretzky et.al., editor, *Proceedings of the 1988 Connectionist Models Summer School*, pages 133–143. Morgan-Kaufman, 1988.

[48] John Moody and Cristian J. Darken. Fast learning in networks of locally-tuned processing units. *Neural Computation*, 1:281 – 294, 1989.

[49] Jooyoung Park and Irwin W. Sandberg. Approximation and radial-basis-function networks. *Neural Computation*, 5(2):305–316, Mar 1993.

[50] K. Pawelzik and H.G. Schuster. Unstable periodic orbits and prediction. *Physical Rev. A*, 43(4):1808–1812, 1991.

[51] John Platt. A resource-allocating network for function interpolation. *Neural Computation*, 3(2):213–225, 1991.

[52] M. J. D. Powell. Radial basis functions for mulivariable interpolation: A review. In J. C. Mason and M. G. Cox, editors, *Algorithms for Approximation*. Clarendon Press, London, 1987.

[53] F. R. Preparata and M. I. Shamos . *Computational Geometry: An Introduction*. Springer-Verlag, 1985.

[54] William H. Press, Brian P. Flannery, Saul A. Teukolsky, and William T. Vetterling. *Numerical Recipes in C*. Cambridge University Press, Cambridge, 1988.

[55] M. B. Priestley. *Spectral Analysis and Time Series*. Academic Press, Inc., London, 1981.

[56] Thorsteinn S. Rögvaldsson. Brownian motion updating of multi-layered perceptrons. In Stan Gielen and Bert Kappen, editors, *Proc. Int. Conf. on Artificial Neural Networks*, pages 527–532, Sep 1993.

[57] O.E. Rössler. An equation for continuous chaos. *Physics Letter*, A(57):397, 1976.

[58] David E. Rumelhart, James L. McLelland, and the PDP Research Group. *Parallel Distributed Processing. Vol. 1-2*. The MIT Press, 1986.

[59] Terence D. Sanger. ? *Neural Computation*, 1990.

[60] Terence D. Sanger. Basis-function trees for approximation in high-dimensional spaces. In *Proceedings of the 1990 Connectionist Models Summer School*. Morgan-Kauffman, 1990.

[61] Tim Sauer, James A. Yorke, and Martin Casdagli. Embedology. *Journal of Statistical Physics*, 65(3/4):579–615, 1991.

[62] David W. Scott. *Multivariate Density Estimation*. John Wiley and Sons, Inc., New York, 1992.

[63] Torsten Söderström and Petre Stoica. *System Identification*. Prentice-Hall, New York, 1989.

[64] K. Stokbro and D. K. Umberger. Forecasting with weighted maps. In M. Casdagli and S. Eubank, editors, *Nonlinear Modelling and Forecasting*, pages 73–94. Addison-Wesley, 1992.

[65] K. Stokbro, D.K. Umberger, and J.A. Hertz. Exploiting neurons with localized receptive fields to learn chaos. *Complex systems*, 4:603, 1990.

[66] M. Stone. Cross-validation, a review. *Math. Operationforsch. Statist. Set. Statist.*, 9:127–139, 1977.

[67] George Sugihara and Robert M. May. Nonlinear forecasting as a way of distinguishing chaos from measurement error in time series. *Nature*, 344:734–741, Apr 1990.

[68] Floris Takens. Detecting strange attractors in turbulence. In D.A. Rand and L.S. Yound, editors, *Dynamical Systems and Turbulence*, pages 366–381. Springer Verlag, Berlin, 1981.

[69] Manoel Fernando Tenorio and Wei-Tsih Lee. Self organizing neural networks for the identification problem. In D. S. Touretzky, editor, *Avances in Neural Information Processing Systems 1*. Morgan Kaufman Publishers, 1989.

[70] H. Tong. *Nonlinear Time Series: A Dynamical System Approach*. Oxford University Press, 1990.

[71] Brent Townshend. Nonlinear prediction of speach signals. In Martin Casdagli and Stephen Eubanks, editors, *Nonlinear Modelling and Forecasting*, pages 433–453. Addison-Wesley, 1992.

[72] A. Weigend, B. Huberman, and D.E. Rumelhart. Prediciting the future: a connectionist approach. *Int. J. Neural Systems*, 1:193–209, 1990.

[73] A. Weigend, B. Huberman, and D.E. Rumelhart. Prediciting sunspots and excange rates with connectionist networks. In M. Casdagli and S. Eubank, editors, *Nonlinear Modelling and Forecasting*, pages 395–432. Addison-Wesley, 1992.

[74] S. T. Welstead. Multilayer feedforward networks can learn strange attractors. In *Proc. Int. Joint Conference of Neural Networks*, pages II139–II144, New York, Jul 1991. IEEE.

[75] B.J. West and H.J. Mackey. Forecasting chaos: A rewiev. *Journal of Scientific and Industrial Reseacrh*, 51:634–643, 1992.

[76] D.M. Wolpert and R.C. Miall. Detecting chaos with neural network. *Proc. R. Soc. London*, B(242):82–86, 1990.

| Method | Prediction time = 6 | | Prediction time = 84 | |
|---|---|---|---|---|
| | $e$ | Reference | $e$ | Reference |
| Global polynomials | 1.1% | Casdagli (89) | | |
| Global rationals | 7.2% | Casdagli (89) | | |
| Multi layer perceptron | 1.0% | Lapedes (87) | 5.0% | Platt (91) |
| Method of analogy | | | 25.1% | Moody (88) |
| Local linear polynomials | 3.3% | Casdagli (89) | 4.5% | Stokbro (92) |
| Local quadratic polynomials | 1.3% | Casdagli (89) | | |
| Standard RBF | 1.1% | Casdagli (89) | 15.8% | Moody88) |
| K-Means RBF | | | 9.3% | Moody (88) |
| Adaptive clustering RBF | | | 7.0% | Platt (91) |
| Hashing RBF | | | 5.0% | Moody (89) |
| Weighted constant map | 6.0% | Stokbro (92) | | |
| Weighted linear map | 1.3% | Stokbro (92) | 3.0% | Stokbro (92) |
| Ridge functions | | | 8.0% | Hartman (91) |
| Coarse coding RBF | | | 5.5% | Deco (93) |

Table 1: Normalized root mean square prediction error for a number of different approximation schemes with two different lengths of the prediction interval. Data is from the Mackey-Glass delay difference equation.

| Data set description | Data type | $m$ | $e_{NL}$ | $e_{AR}$ |
|---|---|---|---|---|
| Mackey-Glass | simulated | 4 | 0.2 | 0.4 |
| Mackey-Glass | simulated | 6 | 0.03 | 0.4 |
| Ikeda map 0% noise | simulated | 5 | <0.02 | 0.9 |
| Ikeda map, 2% noise | simulated | 5 | 0.06 | 0.9 |
| Ikeda map, 20% noise | simulated | 5 | 0.5 | 0.9 |
| Two coupled diodes | lab. data | 7 | 0.3 | 0.9 |
| Four coupled diodes | lab. data | 7 | 0.5 | 0.9 |
| Weak fluid turbulence | lab. data | 20 | 0.01 | 0.4 |
| Strong fluid turbulence | lab. data | 20 | $\approx 0.22$ | 0.16 |
| Flames, non-chaotic | lab. data | 20 | 0.05 | 0.1 |
| Flames, weak-chaotic | lab. data | 20 | 0.12 | 0.25 |
| Flames, strong-chaotic | lab. data | 20 | $\approx 0.7$ | 0.56 |
| Speech | natural data | 20 | 0.2 | 0.3 |
| EEG, resting patient, | natural data | 20 | $\approx 0.7$ | 0.54 |
| EEG, with anaesthesia, | natural data | 20 | $\approx 1.2$ | 0.9 |
| Measles | natural data | 2 | 0.23 | 0.27 |
| Sunspots | natural data | 6 | 0.36 | 0.44 |

Table 2: Table comparing normalised root mean square prediction errors for a global AR model ($e_{AR}$) and the best non-linear model found with local linear AR models ($e_{NL}$). $m$ is the dimension of the coordinate delay vector. From Casdagli (1991).

| Data set | Authors |
|---|---|
| Taylor-Couchette flow | [21] |
| Rayleigh-Bénard convection | [21] |
| Driven semiconductors | [35] |
| Measels | [67], [10] |
| Chickenpox | [67] |
| Marine plankton | [67] |
| Sea surface temperature | [20], [27] |
| Coupled diodes | [10] |
| Fluid turbulence | [10], [20] |
| Flame dynamics | [10] |
| Electroencephallograms | [10] |
| Sunspots | [10], [72] |
| Geology, ground elevation | [12] |
| Computational ecosystem | [72] |
| Double potential well | [27] |

Table 3: Table showing some of the data sets from real processes used in experiments with forecast of chaotic time series.

Figure 1: A) Samples from the logistic map time series plotted in the two dimensional delay coordinate space show the form of the system function $4x(1-x)$. B) The Henon attractor embedded in 3-d space.

Figure 2: The iterates $f^{(r)}$ of a process get more complex as $r$ increases.

Figure 3: The prediction error (ordinate) as a function of prediction time (abscissa) for three different systems and various methods. See the text for details.

Figure 4: A multi layer perceptron with two input variables, three hidden nodes and one output variable implements a function from $\Re^2$ to $\Re$.

Figure 5: The prediction error as a function of the number of training data and the number of parameters. See the text for details

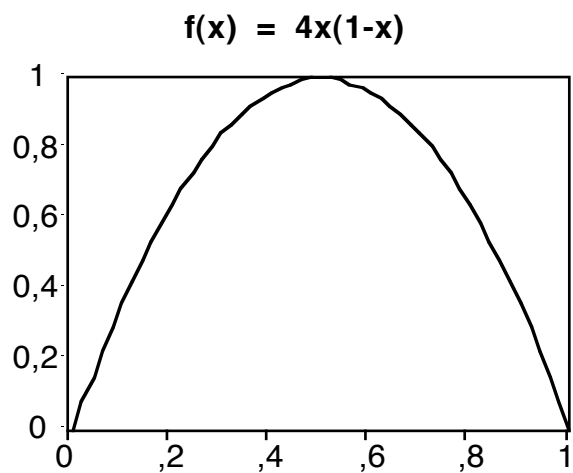
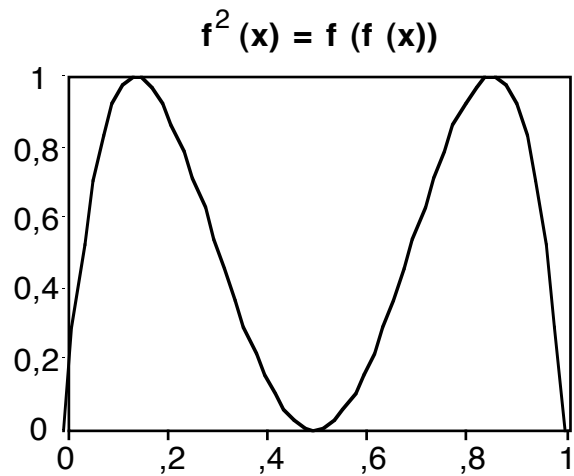
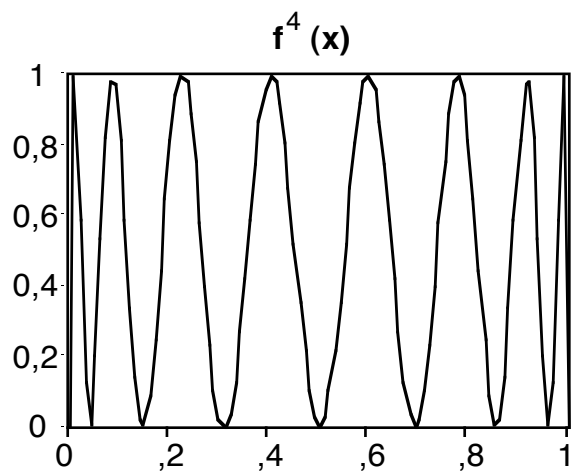
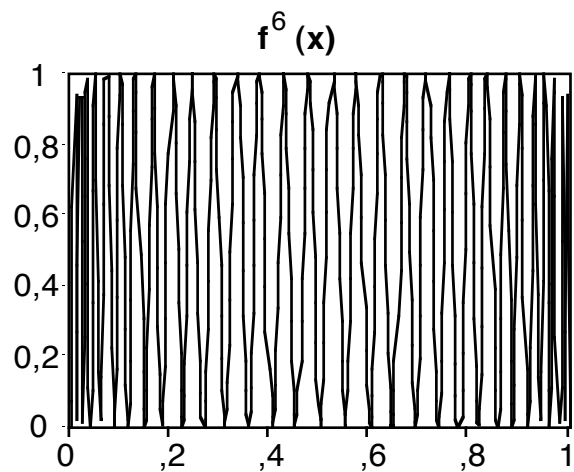

Figure 3
B.Lillekjendlie et. al.
Chaotic time series,
Part II, ...
MIC

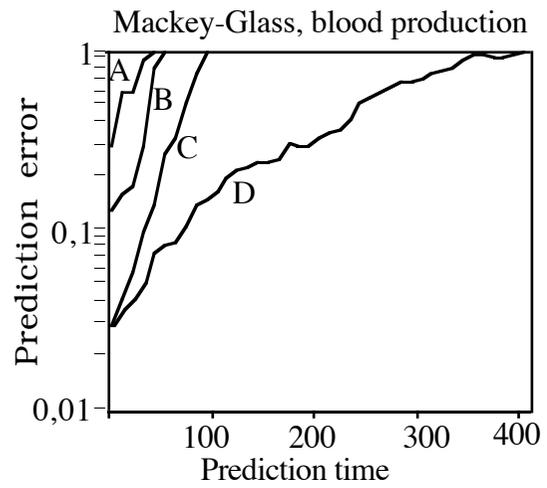 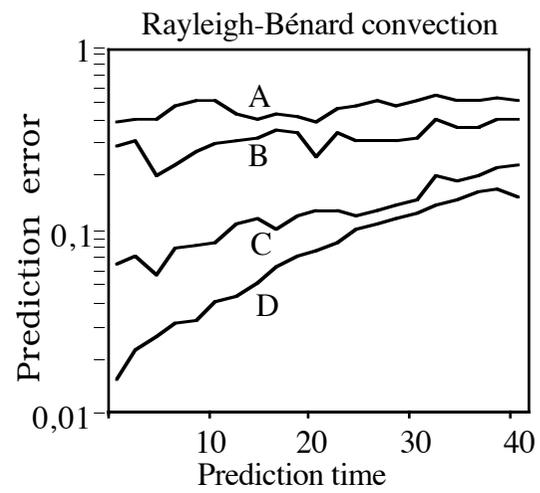 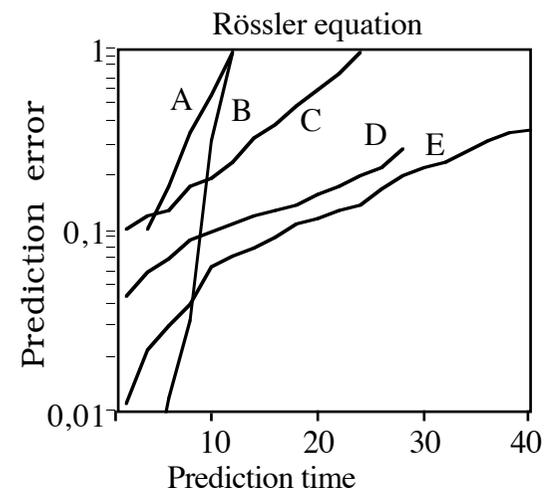

Figure 4
B.Lillekjendlie et. al.
Chaotic time series,
Part II, ...
MIC

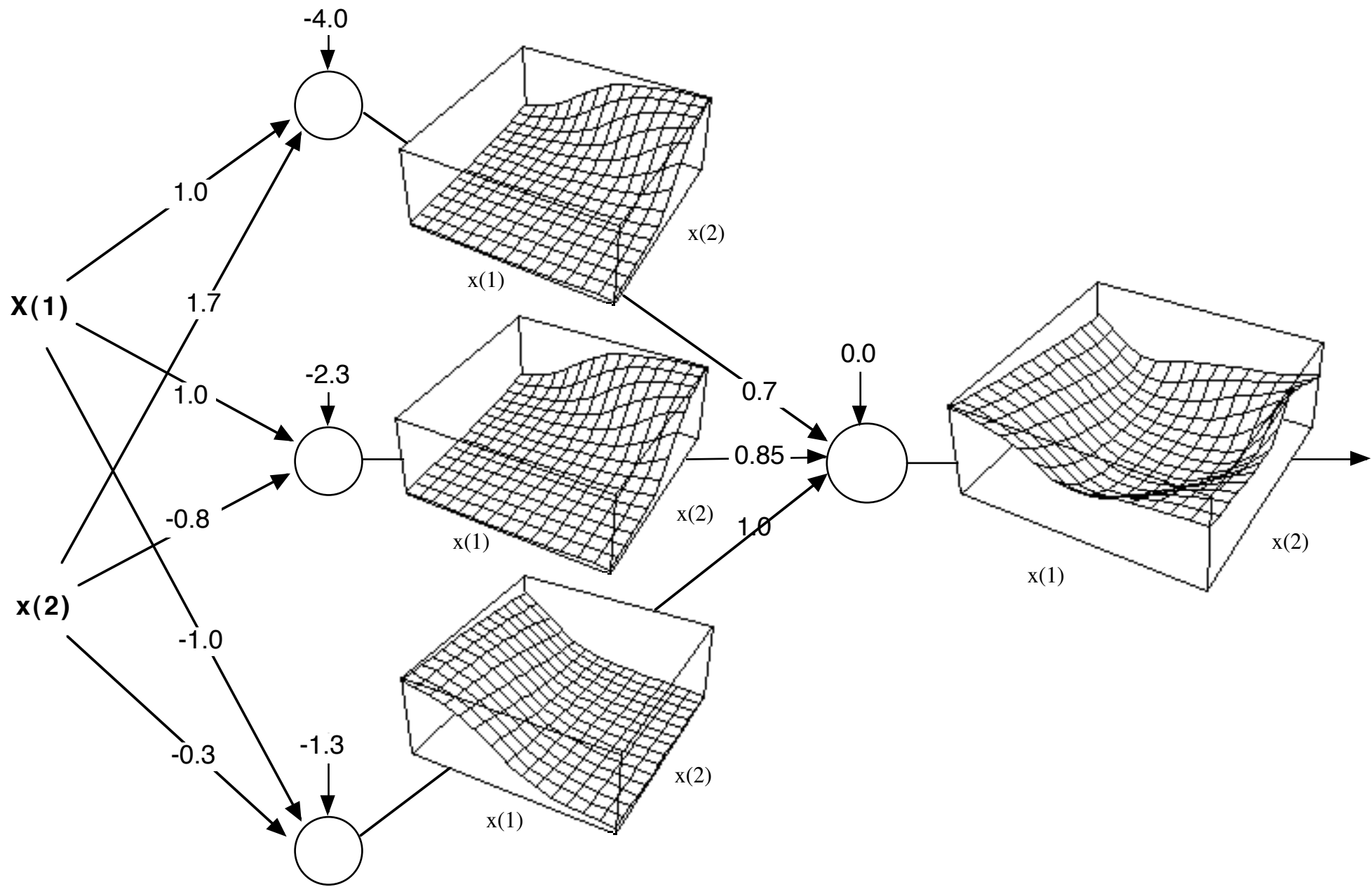

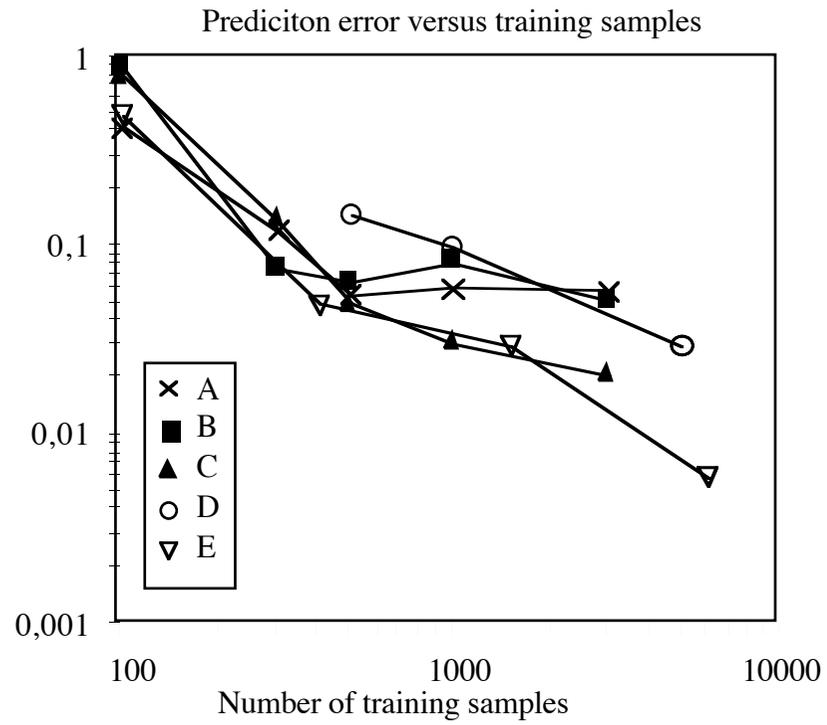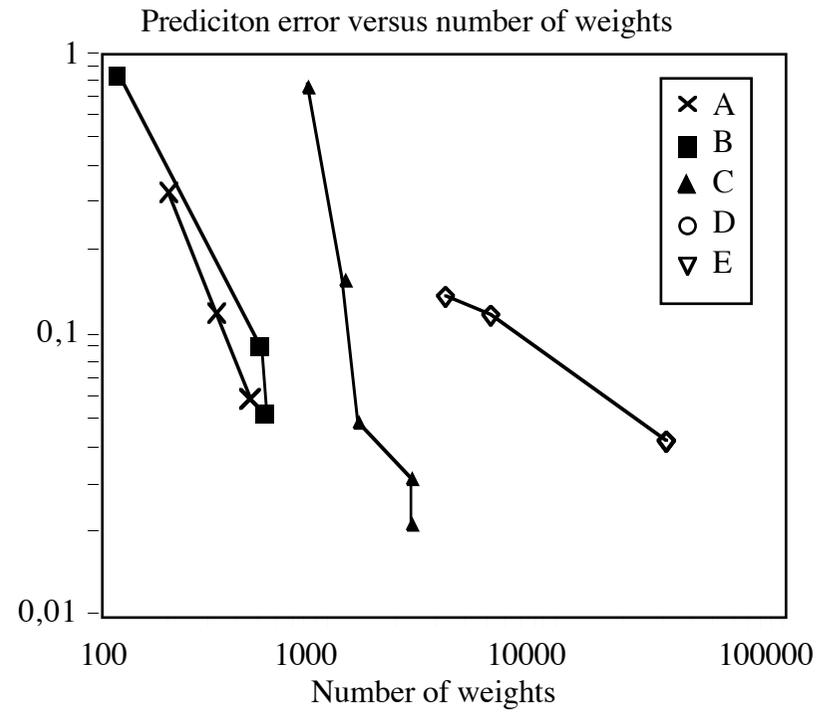

Figur 6
B.Lillekjendlie
MIC artikkel